\documentclass[aps,prb,reprint,superscriptaddress]{revtex4-2} 

\usepackage{graphicx}
\usepackage{amsmath,amssymb,amsfonts}
\usepackage{amsthm}
\usepackage{mathrsfs}
\usepackage{xcolor}
\usepackage{textcomp}
\usepackage{booktabs}
\usepackage{algorithm}
\usepackage{algorithmicx}
\usepackage{algpseudocode}
\usepackage{listings}

\theoremstyle{plain}

\theoremstyle{definition}

\theoremstyle{remark}

\begin{document}

\title{A Depth-Independent Linear Chain Ansatz for Large-Scale Quantum Approximate Optimization}

\author{Zixu Wang}
\affiliation{Department of Materials Science and Engineering, Rensselaer Polytechnic Institute, Troy, NY, 12180, USA}

\author{Jack Mandell}
\affiliation{Department of Mathematical Sciences, Rensselaer Polytechnic Institute, Troy, NY, 12180, USA}

\author{Yangyang Xu}
\affiliation{Department of Mathematical Sciences, Rensselaer Polytechnic Institute, Troy, NY, 12180, USA}

\author{Jian Shi}
\thanks{Correspondence: \href{mailto:shij4@rpi.edu}{shij4@rpi.edu}}
\affiliation{Department of Materials Science and Engineering, Rensselaer Polytechnic Institute, Troy, NY, 12180, USA}
\affiliation{Department of Department of Physics, Applied Physics, and Astronomy, Rensselaer Polytechnic Institute, Troy, NY, 12180, USA}

\date{\today} 

\begin{abstract}
Combinatorial optimization lies at the heart of numerous real-world applications. For a broad category of optimization problems, quantum computing is expected to exhibit quantum speed-up over classic computing. Among various quantum algorithms, the Quantum Approximate Optimization Algorithm (QAOA), as one of variational quantum algorithms, shows promise on demonstrating quantum advantage on noisy intermediate-scale quantum (NISQ) hardware. However, with increasing problem size, the circuit depth demanded by original QAOA scales rapidly and quickly surpasses the threshold at which meaningful results can be obtained. To address this challenge, in this work, we propose a variant of QAOA (termed linear chain QAOA) and demonstrate its advantages over original QAOA on paradigmatic MaxCut problems. In original QAOA, each graph edge is encoded with one entangling gate. In our ansatz, we locate a linear chain from the original MaxCut graph and place entangling gates sequentially along this chain. This linear-chain ansatz is featured by shallow quantum circuits and with the low execution time that scales independently of the problem size. Leveraging this ansatz, we demonstrate an approximation ratio of 0.78 (without post-processing) on non-hardware-native random regular MaxCut instances with 100 vertices in a digital quantum processor using 100 qubits. Our findings offer new insights into the design of hardware-efficient ansatz and point toward a promising route for tackling large-scale combinatorial optimization problems on NISQ devices.

\end{abstract}

\maketitle




\maketitle

\section{Introduction}

\begin{figure*}[t]
\centering
\includegraphics[width=\textwidth]{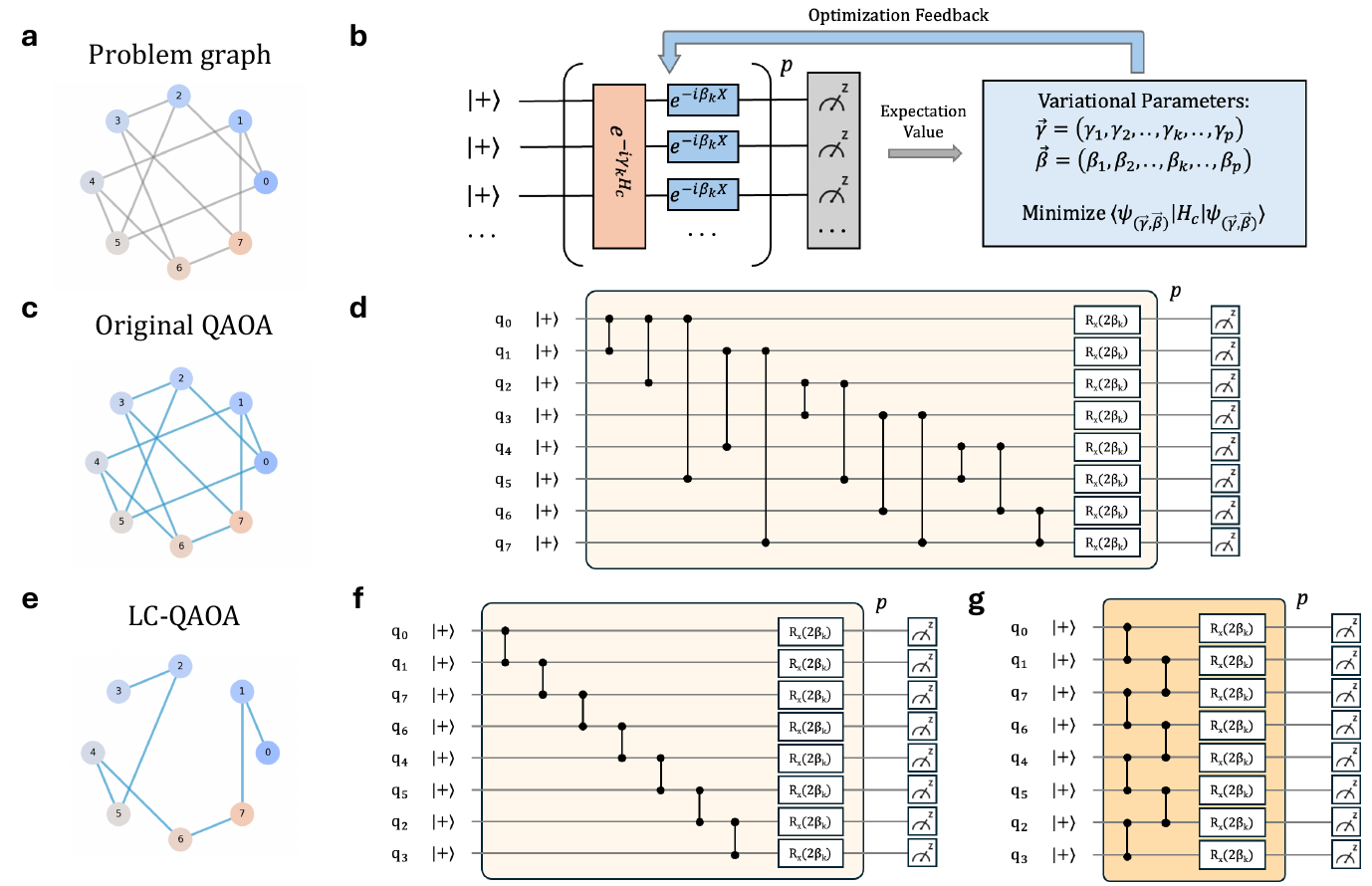}
\caption{
\textbf{Original QAOA and our proposed linear chain QAOA (LC-QAOA).}
\textbf{a,} A random $3$-regular graph with 8 vertices.
\textbf{b,} Quantum-classical hybrid optimization loop of QAOA.
\textbf{c,} In original QAOA, blue edge between vertices is practiced with $R_{zz}$ gate.
\textbf{d,} An original QAOA ansatz circuit with $p$ layers. Here the line connecting two solid dots denotes a $R_{zz}(2\gamma_k)$ gate ($k$ spans from 1 to $p$).
\textbf{e,} In our LC-QAOA, we use a single path in the problem graph with its each blue edge practiced with a $R_{zz}$ gate.
\textbf{f,} Our LC-QAOA ansatz circuit before gate sequence optimization.
\textbf{g,} Our LC-QAOA ansatz circuit with gate sequence optimized in a brick-wall structure. 
}
\end{figure*}

Quantum algorithms are essential in pursuing quantum advantage and addressing real-world problems using quantum computers \cite{childs2010quantum,bharti2022noisy,abbas2024challenges,biamonte2017quantum,montanaro2016quantum,miessen2023quantum}. Many quantum algorithms such as quantum Fourier transform \cite{shor1994algorithms} and quantum adiabatic evolution \cite{farhi2001quantum}, while powerful in the fault-tolerant regime, typically require deep quantum circuits and high-precision gate operation, and are therefore beyond the reach of current hardware \cite{adedoyin2018}. It is thus crucial to design algorithms that are both efficient and compatible with noisy intermediate-scale quantum (NISQ) devices.

Variational Quantum Algorithm (VQA) is a quantum-classical hybrid algorithm featuring a lower depth of circuit and thus suitable for deployment on NISQ. In VQA, the quantum computer performs state evolution and measurement, while the classical computer optimizes parameters based on these measurement outcomes, forming a closed-loop workflow. VQA has been applied in multiple real-world problems in physics, chemistry, etc. and is considered a promising algorithm to demonstrate the quantum advantage in NISQ \cite{kandala2017hardware,cerezo2021,hempel2018quantum}. In VQA, how to build the optimal ansatz, which is a parameterized quantum circuit for quantum state production, is one of the central tasks \cite{sim2019}. Various strategies have been explored for ansatz construction, including problem-inspired ansatzes \cite{taube2006, peruzzo2014}, hardware-efficient designs that respect the device topology \cite{kokail2019, leone2024}, and more universal constructions such as Two-local and brick-wall ansatz. A notable example of ansatz construction is the Quantum Approximate Optimization Algorithm (QAOA), which is a problem-inspired ansatz designed to solve combinatorial optimization problems \cite{farhi2014}. In QAOA, problem to be solved is encoded into a cost Hamiltonian, which is then used as a guideline to introduce entanglement into the ansatz circuit.

Since the proposal of QAOA by Farhi et al \cite{farhi2014} in 2014, many studies have shown its potential to solve various combinatorial optimization problems, such as MaxCut problem (Fig. 1a), traveling salesman problem and spin glass \cite{pelofske2024short,pagano2020quantum,qian2023,egger2021warm,vikstaal2020applying,pelofske2024scaling}. For example, in 2021, Harrigan et al. deployed QAOA on the Google Sycamore quantum processor to solve MaxCut problem and revealed that for hardware-native problems, the approximation ratio (defined to be the cut value computed by quantum computer over true MaxCut value) of QAOA is independent of problem size \cite{harrigan2021}. In 2023, Zhou et al proposed QAOA-in-QAOA to solve arbitrary large-scale MaxCut problems using small quantum machine \cite{zhou2023}. In 2024, Sack et al. applied machine learning for error mitigation and observed a meaningful parameter optimization in $p=$2 QAOA on 40 qubits ($p:$ the number of ansatz layers) \cite{sack2024}. Sachdeva et al. performed QAOA on an IBM quantum processor to solve non-hardware-native unweighted and weighted random $d$-regular graphs (see one example of random $3$-regular in Fig. 1a). In their study, when combined with classical post-processing, true MaxCut can be correctly found in all instances \cite{sachdeva2024}. To enhance the performance of QAOA, researches also modified the QAOA ansatz and developed various QAOA variants, including ma-QAOA \cite{herrman2022}, ADAPT-QAOA \cite{zhu2022}, QAOA+ \cite{chalupnik2022} and LR-QAOA \cite{montanez2024,montanez2025}. These ansatz either show better performance than original QAOA ansatz at the same number of layers or have reduced depth of circuit. Much effort has also been made to optimize hardware compilation, including SWAP network \cite{kivlichan2018}, variation-aware compilation \cite{alam2020} and two-qubit gate number reduction based on edge coloring or depth-first search strategy \cite{majumdar2021}.



However, a bottleneck that remains unresolved for QAOA and its variants is the rapid growth of circuit depth as the problem size increases. For example, solving the MaxCut problem on a random $3$-regular graph with 100 vertices using the original QAOA at $p=$1 requires an circuit execution time of around 160 $\mu s$ on \textit{ibm\_rensselaer}, which is at the same order of magnitude of its relaxation time (around 260 $\mu s$). Such long execution time makes the quantum circuit highly vulnerable to noise arising from limited relaxation times and gate infidelities, thereby severely degrading the optimization outcome. One major contributor to this growth is the SWAP overhead \cite{hirata2011}. Current superconducting quantum computers employ planar coupling maps, such as linear nearest-neighbor, heavy-hex, or square topologies \cite{zhu2025}. However, in practice, the problem of interest are often non-planar. When solving MaxCut problem on non-hardware-native graph, if two qubits lack a direct hardware-native coupling, multiple SWAP gates must be applied to move qubit states until the desired qubits become adjacent, thereby enabling the required entanglement. This process can significantly increase circuit depth \cite{zhu2025}. Even with techniques like SWAP networks \cite{kivlichan2018,weidenfeller2022}, as the problem size grows, the number of SWAP gates can far exceed the number of entangling gates, leading to excessively deep circuits and long execution times. 

To address this issue, in this work, we introduce a linear chain QAOA (LC-QAOA) ansatz and showcase its performance on the MaxCut problem, which is a standard benchmark for evaluating quantum algorithms (see Methods, Section A for details). Our method begins by identifying a long path within the original problem graph and establishing entanglement only between qubits corresponding to adjacent vertices along this path. These qubits are then mapped onto a linear path in the device coupling graph, ensuring hardware-native connectivity and thereby avoiding SWAP overhead. By reorganizing the gates into a bi-layer brick-wall structure, the circuit depth remains independent of the number of qubits, making this approach promising for solving ultra-large problems. Based on our experiments on multiple random $d$-regular graphs executed on quantum hardware, LC-QAOA significantly outperforms the original QAOA. Moreover, the shallow circuit of LC-QAOA allows its ansatz to be extended to high number of layers ($p$) while keeping noise accumulation low. This capability to employ high $p$ ansatz further enhances the achievable approximation ratio (AR).

\section{Experimental Design}

In solving combinatorial optimization problems with VQA, the first step is to reformulate the problem as a cost function, where the task is transformed to finding its minimum. This cost function is then mapped onto a Hamiltonian, typically through a Quadratic Unconstrained Binary Optimization(QUBO) to Ising transformation, such that the ground state of the Hamiltonian encodes the optimal solution. In practice, the search for the ground-state eigenvalue and eigenvector is carried out through a variational procedure. An ansatz (Fig. 1b) is first employed to generate a trial quantum state parameterized by a set of classical variables. The quantum computer is then used to evaluate the expectation value of the Hamiltonian with respect to this trial state. Based on the measurement outcomes, a classical optimizer updates the ansatz parameters for new trial state generation. The quantum-classical process is performed iteratively to minimize the expectation value. The final expectation value serves as an approximation to the ground state eigenvalue of the Hamiltonian, with the corresponding state approximating the ground-state eigenvector thereby yielding the solution to the original optimization problem. A detailed description of how to construct the MaxCut cost function from the problem and encode the cost function into the QAOA Hamiltonian framework is provided in the Methods (Section A).

QAOA is regarded as a type of VQA with a problem-inspired ansatz to solve combinatorial optimization problems, where the Hamiltonian encoding the optimization problem is directly involved in the ansatz construction (Fig. 1b). 
The evolved state under QAOA ansatz can be expressed as below:
\begin{equation}
|\psi_p(\gamma,\beta)\rangle = e^{-i\beta_p \hat{H}_M}\, e^{-i\gamma_p \hat{H}_C} \cdots e^{-i\beta_1 \hat{H}_M}\, e^{-i\gamma_1 \hat{H}_C} |+\rangle^{\otimes N}
\end{equation}
where $|\psi_p(\gamma,\beta)\rangle$ is the trial state determined by a set of parameters $(\gamma_1,\beta_1,...,\gamma_p,\beta_p)$. $\gamma$ and $\beta$ are the variational parameters to optimize. $\hat{H}_C$ is the cost Hamiltonian, converted from the problem-defined cost function, implemented through a series of two-qubit $R_{zz}(2\gamma)$ gates in the quantum circuit. $\hat{H}_M$ is the mixing Hamiltonian, corresponding to the time evolution under $H_M = \sum_i X_i$, realized as a set of parallel single-qubit rotations $R_x(2\beta)$. 
As the number of layers $p$ increases, more parameters are involved and the expressibility of the ansatz improves. This enlarged variational space increases the likelihood of containing states closer to the theoretical optimum.

Based on the ansatz formulation, the original QAOA can be visualized in Fig. 1d: for each edge in the problem graph (Fig. 1c), an $R_{ZZ}$ entangling gate is inserted between the corresponding qubits. On superconducting quantum processors with limited qubit connectivity, additional SWAP gates are often required to bring non-adjacent qubits together. As the problem size and the number of qubits increase, the overhead from these SWAP operations grows rapidly, thereby hindering the scalability of QAOA and limiting its potential to demonstrate quantum advantage.

In our proposed LC-QAOA, first, we identify a long path (Fig. 1e) in the problem graph using a depth-first search (DFS) algorithm (see Methods Section B.1 for details and discussion). Next, $R_{ZZ}$ gates are applied along this chain to introduce entanglement, as shown in Fig. 1f. Since the two-qubit gates are placed along a linear path, non-adjacent $R_{ZZ}$ gates that do not have shared qubits can be executed independently (Fig. 1f). To further improve parallelization, the adjacent gates on the chain are arranged in a two-layer brick-wall structure (Fig. 1g): half of them are executed in the first layer and the remaining half in the second, both simultaneously. This brick-wall arrangement of $R_{ZZ}$ gates defines the cost layer, followed by a mixer layer composed of $R_X$ rotations applied to all qubits. Together, these cost and mixer layers form a single layer of the LC-QAOA ansatz (Fig. 1g). By this, we reduce the execution time of the quantum circuit by reducing the number of entangling operations and eliminating the SWAP overhead.

\section{Results and Discussions}

\begin{figure}[t]
\centering
\includegraphics[width=\linewidth]{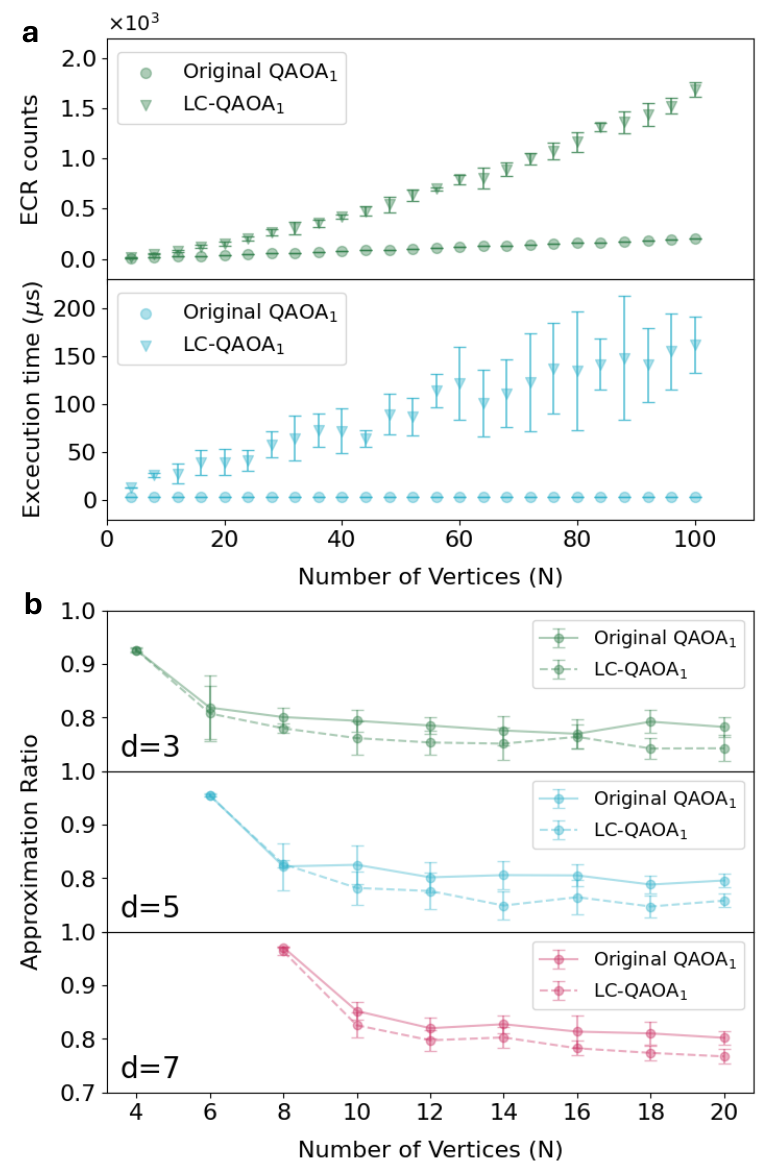}
\caption{
\textbf{Circuit complexity (\texttt{ibm\_rensselaer}) and approximation ratio (noiseless \texttt{Aer Simulator}).}
Here, QAOA$_p$ refers to QAOA with $p$ layers. Each data point represents the average result over 10 randomly generated $d$-regular graph instances.
\textbf{a}, Number of two-qubit echoed cross-resonance (ECR) gate (top panel), and execution time (bottom panel) of original QAOA$_1$ and LC-QAOA$_1$ ansatz circuit at different number of vertices $N$ (on \texttt{ibm\_rensselaer}). Error bar indicate $\pm2\sigma$.
\textbf{b}, Approximation ratio of original QAOA$_1$ and LC-QAOA$_1$ on random $d$-regular graph at different number of vertices (on noiseless \texttt{Aer Simulator}). Error bar indicate $\pm\sigma$.
}
\end{figure}

\begin{figure*}[t]
\centering
\includegraphics[width=\textwidth]{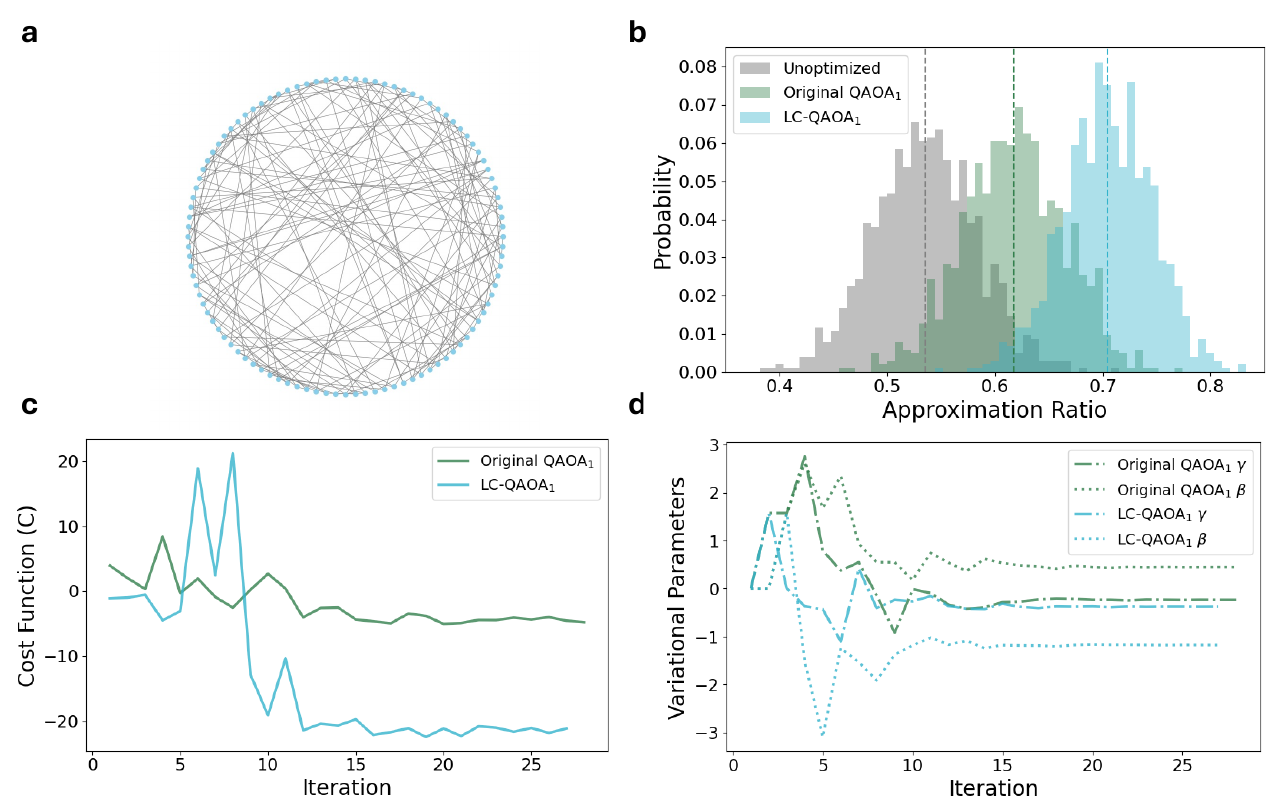}
\caption{
\textbf{Hardware implementation of solving MaxCut problem on random $3$-regular graph with 100 vertices (executed on \textit{ibm\_kingston}).}
\textbf{a,} The 100-vertices problem instance.
\textbf{b,} Probability distribution of non-post-processed approximation ratio of unoptimized state, original QAOA$_1$ and LC-QAOA$_1$. Both QAOA are initialized from $|+\rangle^{\otimes N}$ state. All the states are sampled for 1024 times. Vertical dash line denotes the mean approximation ratio.
\textbf{c,} Cost function value ($C$) of original QAOA$_1$ and LC-QAOA$_1$ with respect to number of iterations.
\textbf{d,} The evolution of variational parameters with respect to number of iterations.
}
\end{figure*}

\subsection{Circuit complexity and approximation ratio}


By adding $R_{ZZ}$ gates along a chain (Fig. 1f) and arranging them in a brick-wall structure (Fig. 1g), the key advantage of LC-QAOA ansatz is that its circuit depth remains independent of the size of the problem, enabling efficient scaling to larger problem instance on current quantum hardware. A direct comparison on the counts of native two-qubit echoed cross-resonance (ECR) gates \cite{ecrgate2025} and execution time of a $p=1$ circuit between original QAOA ansatz and our LC-QAOA ansatz is shown in Fig. 2a. As the number of vertices increases, the two-qubit gate count of the original QAOA ansatz grows rapidly, primarily due to SWAP overhead. In contrast, the number of two-qubit gates in the LC-QAOA ansatz increases linearly with the number of vertices, with a slope of one. Similarly, the circuit execution time increases rapidly for the original QAOA ansatz, while it remains constant for our LC-QAOA ansatz.

A natural concern regarding LC-QAOA is that, compared with the original QAOA, its ansatz employs fewer entangling gates, which may reduce its expressibility and degrade the performance of the algorithm. To investigate this, we compared the mean approximation ratios of original QAOA$_1$ and LC-QAOA$_1$ (subscript denotes the number of layers) on multiple random $d$-regular MaxCut instances with $d = 3, 5, 7$ using the noiseless Aer Simulator \cite{aer2025} in Fig. 2b. The mean approximation ratio, defined as the mean cut value calculated from the sampled bitstring set divided by the true MaxCut value, serves as a key performance metric. The results show that while LC-QAOA indeed achieves a slightly lower approximation ratio than the original QAOA at the same number of layers $p$, the gap does not appear to widen with increasing problem size. Although LC-QAOA performs slightly worse than the original QAOA in noiseless simulator, it is our hope that its shallow circuit may bring a net gain on the computing performance if executed on quantum hardware.

\subsection{Experiment deployment on quantum hardware}

We demonstrate our LC-QAOA on IBM’s superconducting quantum processors, and as an example present the MaxCut result for a non-hardware-native random $3$-regular graph with 100 vertices (Fig. 3a). We employed the gradient-free \texttt{COBYLA} optimizer from the \texttt{SciPy} package as the classical optimizer, owing to its fast convergence \cite{virtanen2020}. The optimization was terminated once predefined convergence criteria (tol = $10^{-3}$) is met. Most results in this work are obtained without error mitigation or post-processing unless otherwise noted. Further details of the hardware deployment can be found in Methods (Section C). Additional results for solving MaxCut on weighted and unweighted random $d$-regular graphs of various sizes and degrees executed across different quantum processors are provided in the Supplementary Information (Section A, B, C and D, Table S1, Table S2, Table S3, Fig. S1, S2 and S3).

Fig. 3b presents the probability distribution of approximation ratio obtained from LC-QAOA$_1$ and original QAOA$_1$ (QAOA$_p$ refers to QAOA with $p$ layers). The distribution obtained from the original QAOA is shifted to higher value relative to the unoptimized bitstring set, achieving a mean approximation ratio of 0.62. This demonstrates that even for a large graph with 100 vertices, the original QAOA still yields meaningful results on a IBM’s Heron R2 quantum processor (\texttt{ibm\_kingston}). By contrast, LC-QAOA yields a distribution with a much higher mean approximation ratio of 0.72, demonstrating clear improvement over the original QAOA. This is likely due to the shallower depth of its circuit.

 Fig. 3c shows the cost function versus the number of iterations for original QAOA$_1$ and LC-QAOA$_1$. The cost function asymptotically approaches a stable value as the number of iterations increases, indicating the gradual improvement of the trial state. Compared with the original QAOA$_1$, LC-QAOA$_1$ achieves a lower estimated cost function values on the quantum hardware, reflecting better performance. Fig. 3d shows the evolution of the variational parameters ($\gamma$ and $\beta$) of original QAOA$_1$ and LC-QAOA$_1$ throughout the optimization process.

\begin{figure}[b]
\centering
\includegraphics[width=0.8\linewidth]{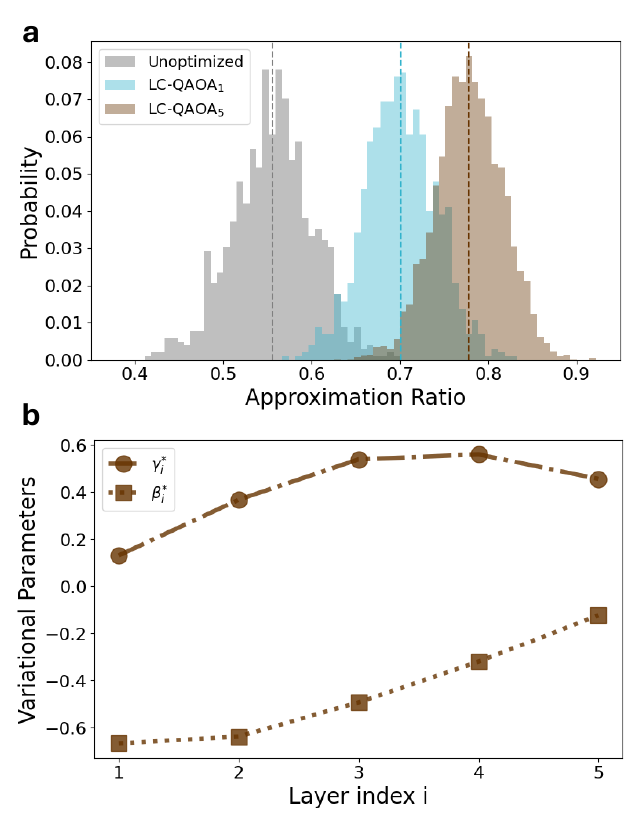}
\caption{
\textbf{Performance of high-$p$ LC-QAOA.}
LC-QAOA$_1$ and LC-QAOA$_5$ are used to solve random $3$-regular graph with 100 vertices MaxCut instance on \textit{ibm\_kingston}. For both LC-QAOA$_1$ and LC-QAOA$_5$, fractional gates (more on section D) are used. All the states are sampled for 4096 times.
\textbf{a,} Probability distribution of non-post-processed approximation ratio of unoptimized state, LC-QAOA$_1$ and LC-QAOA$_5$. Vertical dash line denotes the mean approximation ratio.
\textbf{b,} Optimized variational parameter $\gamma$ and $\beta$ of LC-QAOA$_5$ in different layers.
}
\end{figure}

\subsection{Performance of high $p$ ansatz}

In theory, the expressibility of the ansatz improves as $p$ increases, offering the potential for a better final state \cite{farhi2014,farhi2020quantum,basso2021quantum,boulebnane2025evidence,wurtz2021maxcut}. In practice, however, the original QAOA suffers from noise accumulation as circuit depth grows, which quickly offsets the advantages of higher expressibility. By contrast, it is expected the short execution time of LC-QAOA ensures that the performance gains from larger-$p$ ansatz can outweigh noise-induced degradation. One challenge in exploiting high-$p$ circuits lies in parameter optimization: as $p$ increases, the number of parameters grows and the cost-function landscape becomes more intricate, making optimization increasingly difficult for classical optimizers. To mitigate this issue, we adopt a heuristic parameter initialization strategy proposed by Zhou et al. \cite{zhou2020}. Their analysis of multiple QAOA instances shows that the optimal parameters across different layers are not arbitrary but exhibit discernible patterns. This pattern allows us to guess the optimal variational parameter of QAOA$_p$ based on the optimized variational parameter of QAOA$_{p-1}$ and use the guessed value as an initial point to speed up parameter optimization. In this work, we used the FOURIER heuristics they proposed for high $p$ circuit initialization.

Fig. 4a presents the probability distribution of approximation ratio of LC-QAOA$_1$ and LC-QAOA$_5$ on a MaxCut instance. Compared to the $p=1$ case ($AR$=0.70), the $p=5$ ansatz achieves a mean approximation ratio of 0.78 and a best approximation ratio of 0.92, without any error mitigation or post-processing. These experimental results further confirm the potential of increasing $p$ in QAOA to obtain improved performance, validating the effectiveness of the heuristic parameter initialization. Fig. 4b presents the final ansatz parameters obtained from the LC-QAOA$_5$ optimization.

\subsection{Fractional gate}
Recently IBM developed a new quantum gate, called fractional gate in their Heron processors, including one-qubit $R_X$ and two-qubit $R_{zz}$ gate. These hardware-native gates can be used to replace their previously synthesized counterparts from the conventional digital gate set (Fig. 5a). The use of fractional gates substantially reduces both the depth and duration of circuits, which is especially important for high-$p$ circuits \cite{fractionalgate2025}. Here, we demonstrate the effect of fractional gate using LC-QAOA$_5$. Figure 5b compares the approximation ratio distributions of circuits with and without fractional gates on the same MaxCut instance. It shows that hardware native $R_{zz}$ gate and $R_X$ gate outperform the synthesized case slightly.


\begin{figure}[t]
\centering
\includegraphics[width=0.9\linewidth]{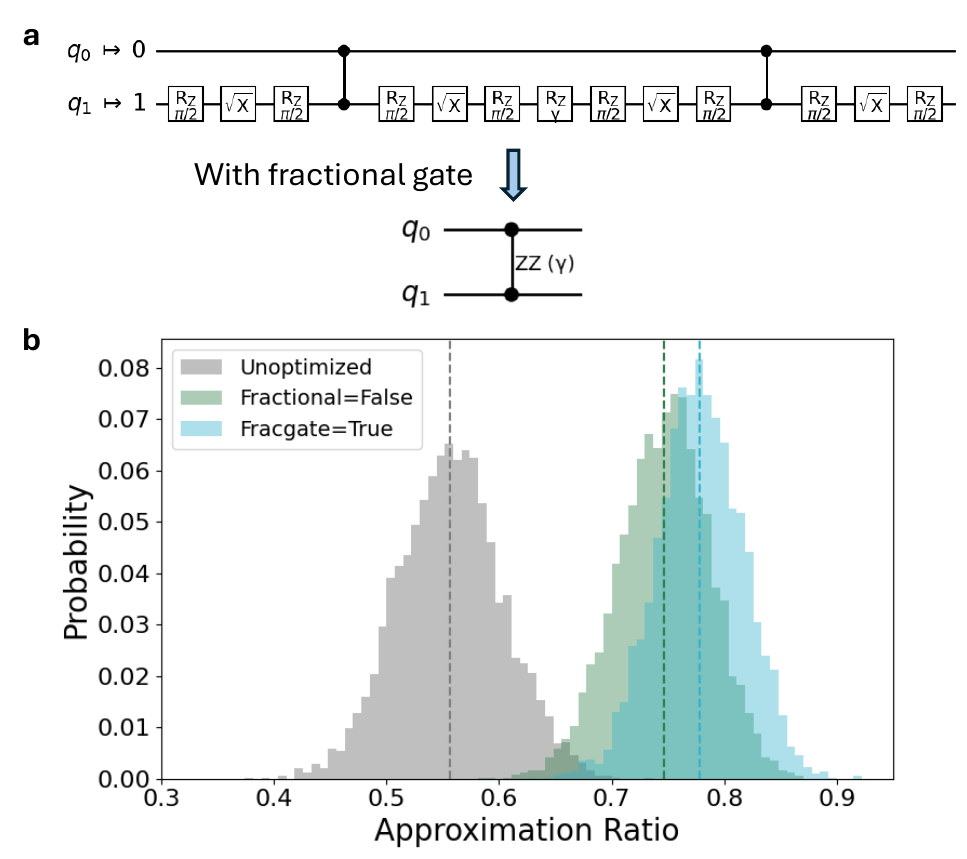}
\caption{
\textbf{The effect of fractional gates in LC-QAOA$_5$.}
\textbf{a,} $R_{zz}$ gate hardware implementation with and without fractional gate feature. Without fractional gate feature, one $R_{zz}$ gate is synthesized by two CZ gates and a series of single-qubit gates. With fractional gate feature, $R_{zz}$ gate is hardware native.
\textbf{b,} Probability distribution of approximation ratio of a random $3$-regular graph with 100 vertices MaxCut instance solved by LC-QAOA$_5$ with and without fractional gate feature executed on \textit{ibm\_kingston}. All the states are sampled for 4096 times.
}
\end{figure}

\subsection{Classical bit-flip post-processing}
Bit-flip post-processing is a well-established technique to improve solution quality in combinatorial optimization problems \cite{sciorilli2025,sachdeva2024,boulebnane2020}. This method is a heuristic local search procedure designed to escape suboptimal solutions by iteratively seeking local improvements. In this work, we apply the bit-flip post-processing to the bitstring set sampled from the final state of LC-QAOA$_2$ to further improve its quality. Specifically, each bit of a sampled bitstring is flipped in turn, and the cut value of the modified bitstring is evaluated. If the cut value increases, the new bitstring is retained and the procedure continues; otherwise, the flip is discarded and the algorithm proceeds to the next bit. Fig. 6 presents a probability distribution of approximation ratio before and after bit-flip post-processing. The results show that the bit-flip step significantly improves the bitstring set quality, raising the mean approximation ratio from 0.72 to 0.95 and successfully recovering the true MaxCut solution (maximum $AR$=1). We further applied this post-processing strategy to multiple random $d$-regular MaxCut instances solved by LC-QAOA, all of which achieved true MaxCut results after post-processing, as shown in the Supplementary Information (Section E).

\begin{figure}[t]
\centering
\includegraphics[width=0.9\linewidth]{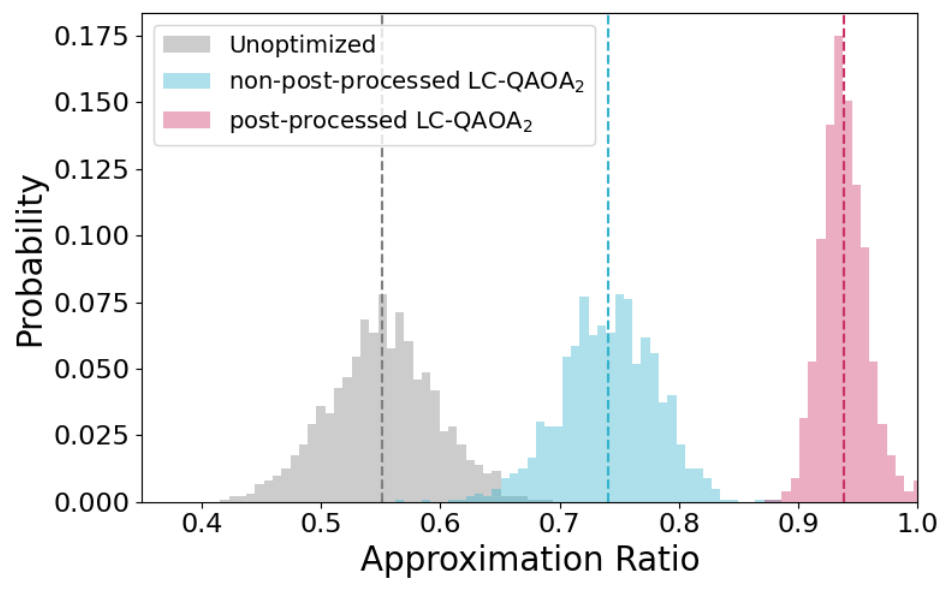}
\caption{
\textbf{Classical bit-flip post-processing.}
Probability distribution of approximation ratio of unoptimized state, non-post-processed LC-QAOA$_2$ and post-processed LC-QAOA$_2$. All the states are sampled for 1024 times.
}
\end{figure}

\section{Conclusion}
In this work, we propose a new QAOA ansatz that is featured by a linear topology in a random regular graph. This scheme eliminates the need for SWAP operations and significantly reduces circuit depth. The shallow circuit allows to tackle problem of large number of vertices with high-$p$ ansatz with limited impact from circuit noise. Using this linear chain ansatz, we solve the MaxCut problem on multiple random $d$-regular graphs with up to 120 vertices on IBM's quantum processor. Without error mitigation or post-processing, the mean approximation ratio reaches 0.70 for the $p=1$ circuit and 0.78 for the $p=5$ circuit, which, to the best of our knowledge, represents the state-of-the-art results among all QAOA variants. It is noted that although higher approximation ratios have been reported, some of these results are only demonstrated in specific instances \cite{wang2025,pelofske2024scaling}, while others rely on parameter optimization performed on local classical simulators (optimized circuits are then executed on quantum hardware for bitstring sampling) \cite{sciorilli2025,shaydulin2023}. By contrast, our ansatz exhibits consistent performance across multiple random instances, with quantum hardware engaged throughout the entire parameter optimization process.

An important question is whether our results can ultimately lead to quantum advantage on NISQ devices. We note that, for classical exact solvers, execution time grows rapidly with both the number of vertices and the degree of the graph. In our LC-QAOA the circuit depth does not scale with the problem size. Although the instances used in our current demonstrations limited by the number of qubits available from existing quantum hardware are easily solvable by classical solvers, it is plausible that quantum advantage may emerge if quantum hardware with much higher number of qubits becomes available. 

\section{Methods}

\subsection{MaxCut Problem}
MaxCut is an important problem in combinatorial optimization and graph theory. The problem is defined on an undirected graph $G=(V,E)$, where $V$ is the set of vertices and $E$ is the set of edges. For weighted MaxCut problem, a weight $\omega$ will be assigned to each edge (for unweighted MaxCut problem, all $\omega$ equals 1). The task is to partition the vertex set into two disjoint subsets such that the number (for unweighted MaxCut problem) or weight (for weighted MaxCut problem) of edges crossing between the two subsets is maximized, respectively.
Below are detailed steps about how to transform MaxCut problem to cost Hamiltonian for QAOA algorithm. 
First, we have,
\begin{equation}
\delta_{x_i,x_j} = x_i + x_j - 2x_i x_j
\end{equation}
where $x_i, x_j \in \{0,1\}$, represent the group assignments of the vertices. The quantity $\delta_{x_i,x_j}$ equals 1 if $x_i$ and $x_j$ are different and 0 otherwise.
Then we define a cost function $C(x)$ over the whole graph, where $x \in \{0,1\}^{|V|}$.
\begin{equation}
C(x) = \sum_{x_i x_j \in E} \omega_{i,j}\left( x_i + x_j - 2x_i x_j \right)
\end{equation}
The cost function $C(x)$ evaluates the cut value associated with a given binary assignment $x$. The MaxCut problem can thus be formulated as a Quadratic Unconstrained Binary Optimization (QUBO) problem, where maximizing $C(x)$ is equivalent to minimizing $-C(x)$. For quantum algorithms, the cost function is further encoded into a Hamiltonian, 
whose ground state corresponds to the optimal solution.
\begin{equation}
-C(x) = x^{T} Q x
\end{equation}
where $Q$ is a real symmetric matrix. We can further decompose $Q$ into linear combination of Ising Hamiltonian and identity operator as below:
\begin{equation}
Q = H_{\text{Ising}} + c \cdot I
\end{equation}
where $H_\text{Ising}$ denotes a linear combination of Ising Hamiltonian, $c$ is a constant, and $I$ is the identity operator.
QAOA is then used to search for the state with lowest energy for $H_\text{Ising}$.
The final state is then sampled and output a bitstring set. Cut value of every bitstring is calculated. The bitstring with highest cut value is selected as the final solution.

\subsection{Numerical Details}

\subsubsection{Longest path search}
In this research, a greedy Depth-First Search (DFS) method is used to find the longest path in the graph to build entanglement. It is worth noticing that finding the longest path in a graph itself is an NP-hard problem. Heuristic method is used for this task, which provides no guarantee of achieving the maximum chain length in the graph. At the same time, for a random $d$-regular graph, a Hamiltonian path (a path that visits each vertex exactly once) may not exist. As a result of both, some vertices may be excluded from the linear path, particularly when the graph size is large and degree of the graph is low. To further explore how this will affect efficiency of LC-QAOA, we run experiments on noiseless \texttt{Aer Simulator} to reveal the relationship between length of the linear chain for a fixed number of total vertices in the graph and approximation ratio in LC-QAOA$_1$. With shorter linear chain, the LC-QAOA ansatz remain functional with lower approximation ratio. Data is shown in Supplementary Information (Section F, Fig. S4).

\subsubsection{Quantum simulator}
In this work, noiseless Aer Simulator from \texttt{qiskit\_aer} is used for approximation ratio simulation in Fig. 2a up to 20 qubits.

\subsubsection{Classical bit-flip post-processing}
As mentioned in the main text, classical bit-flip post-processing is done to further improve the quality of bitstring set sampled from the final state. The procedure terminates once a full traversal of the bitstring yields no further improvement. A flowchart of this process is presented in Fig. 7.

\begin{figure}[t]
\centering
\includegraphics[width=\linewidth]{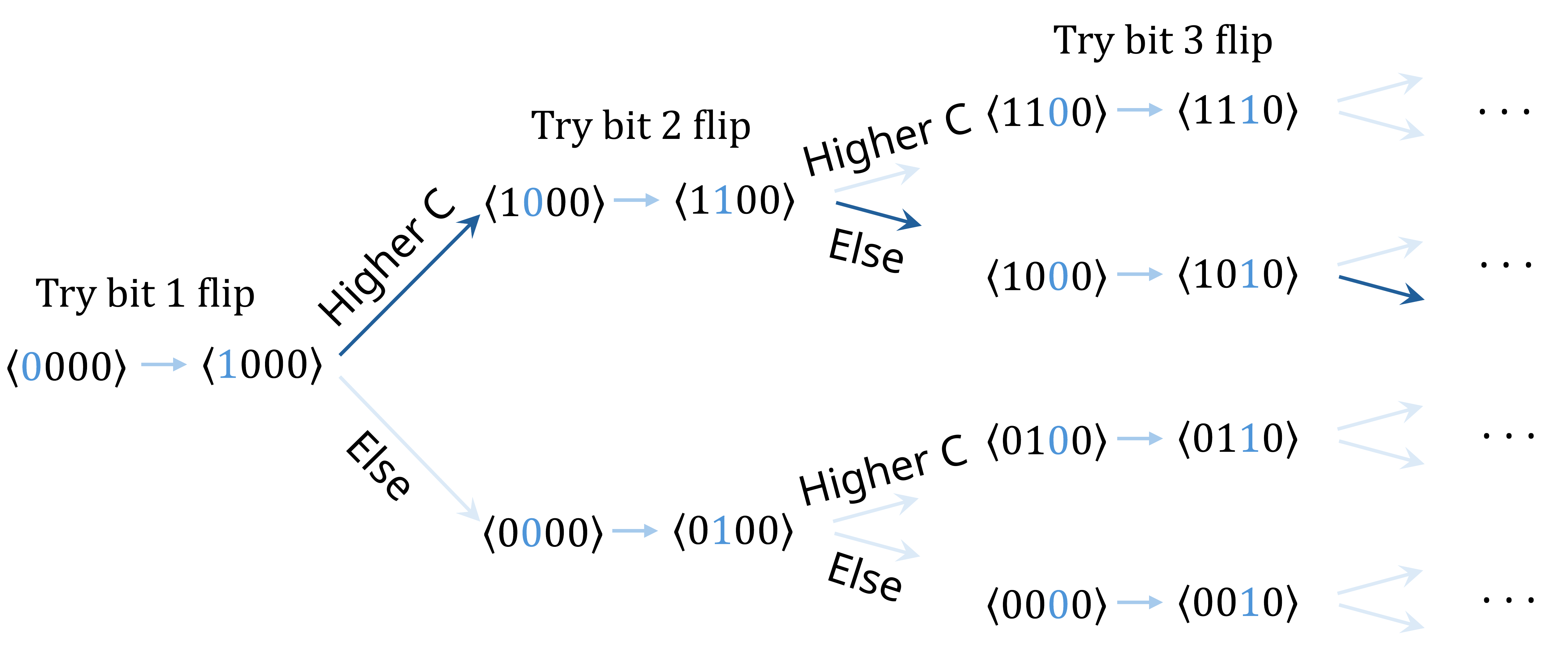}
\caption{
\textbf{Schematic illustration of the bit-flip post-processing procedure.}
Starting from the initial state $\langle 000\cdots0 \rangle$ (here we use a four-bit state as one example), individual bits are sequentially flipped and the cost function $C$ is evaluated after each flip. If the flip yields a higher cost ($Higher\text{ }C$), the new state is accepted; otherwise ($Else$), the state reverts to the previous configuration. The dark blue path highlights one possible trajectory of accepted flips.
}
\end{figure}

\subsubsection{Exact MaxCut calculation}
In this work, exact MaxCut value of the MaxCut instances are calculated by \texttt{Gurobi} solver \cite{gurobi2024}.


\subsection{Experiment Details}
\begin{figure*}[t]
\centering
\includegraphics[width=0.8\textwidth]{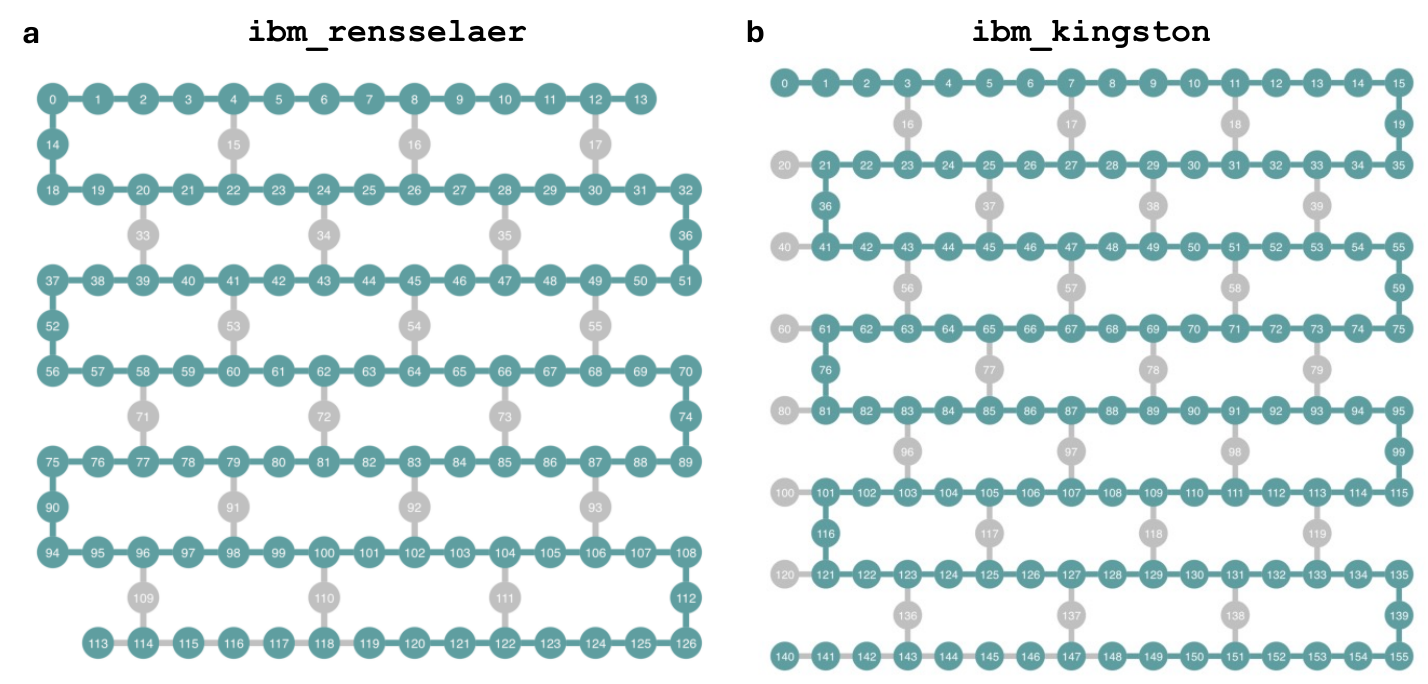}
\caption{
\textbf{Coupling map and our linear chain selection in quantum hardware.}
Dots and lines denote qubits and couplings respectively. Green dots and lines denote qubits and couplings used in linear chain ansatz.
\textbf{a,} \textit{ibm\_rensselaer}.
\textbf{b,} \textit{ibm\_kingston}.
}
\end{figure*}
\subsubsection{Choice of instances}
In this work, all the MaxCut instances are generated using the Python package \texttt{NetworkX} \cite{hagberg2008}.

\subsubsection{Quantum processors}
Two IBM's superconducting quantum processing units (QPUs), \textit{ibm\_rensselaer} and \textit{ibm\_kingston} are used in this work. \textit{ibm\_rensselaer} is an Eagle R3 processor with 128 qubits, equipped with hardware-native $ecr, id, R_z, sx, x$ gates (Fig. 8a). \textit{ibm\_kingston} is a Heron R2 processor with 156 qubits equipped with hardware-native $cz, id, R_x, R_z, R_{zz}, sx, x$ gates (Fig. 8b). Both QPUs adopt heavy-hex coupling map.

\subsubsection{Circuit compilation}
For original QAOA, the circuit is transpiled with optimization level = 1 without defining the initial layout.
For LC-QAOA, the virtual qubits are mapped to a physical chain that connects the same number of physical qubits (Fig. 8). Then the circuit is transpiled with optimization level = 1 with the predefined initial layout.

\subsubsection{Circuit execution}
Two primitives from \texttt{qiskit\_ibm\_runtime} are used in this work. \texttt{EstimatorV2} \cite{Estimator2025} is used to estimate the expectation value of the Hamiltonian with respect to trail state and \texttt{SamplerV2} \cite{sampler2025} is used to sample the final state and return bitstring set. For most of the experiments, we adopted 1024 readout shots for estimator and sampler, unless otherwise noted.

\subsubsection{Classical optimizer}
In VQA, classical optimizers are used for parameters optimization. Multiple optimizers have been adopted, including COBYLA, SPSA, Nelder-Mead \cite{kostas2024}. In this work, gradient-free \texttt{COBYLA} optimizer is used for its fast convergence speed. The optimizer is imported from \texttt{SciPy} package \cite{virtanen2020}. The initial value is set to be 0 for all parameters (i.e. $\beta$ and $\gamma$). For all instance, this optimization is terminated when convergence criteria is met. The convergence tolerance is set to be tol = $1\times 10^{-3}$ without limiting the number of iteration.

\section{Acknowledgments}
This work was supported by IBM through the IBM-Rensselaer Future of Computing Research Collaboration. We thank the Quantum Algorithm Engineering team at IBM for valuable discussions and technical support scaling up quantum workflows.
We would like to express our sincere gratitude to Haimeng Zhang, Nate Earnest-Noble, Alexis Lighten and Jie Chen from IBM for their valuable suggestions and insightful discussions throughout this work. We also thank Michael Sofka from FOCI RPI for his technical help in using the quantum hardware. J.S. acknowledges support from Simons Foundation.

\section{Data and Code statement}
Raw data and codes for this work are available from the corresponding author upon request.

\newpage
\bibliographystyle{apsrev4-2}
\bibliography{Quantum_computing}


\onecolumngrid
\newpage
\section{Supplementary Information}
\renewcommand{\thepage}{S\arabic{page}}
\setcounter{page}{1}
\subsection{Performance of LC-QAOA$_1$ and LC-QAOA$_2$ to solve random $3$-regular MaxCut instances with 100 vertices}
Here we show the result of LC-QAOA on multiple random $3$-regular graphs with 100 vertices MaxCut instances to further validate its effectiveness (Table. 1 and Fig. 9). Two IBM's QPUs (\textit{ibm\_rensselaer} and \textit{ibm\_kingston}) are used to execute the algorithm.

Three observations are made based on these results.
First, different MaxCut instances of random $3$-regular graph with 100 vertices shows similar mean approximation ratio when number of layers $p$ and QPU are the same.
Second, for most of the instances, LC-QAOA$_2$ shows higher approximation ratio than LC-QAOA$_1$ when executed on same QPUs, likely due to the better expressibility at higher number of layers. Note that heuristic parameter initialization is not used for $p$=2 results.
Third, for the same MaxCut instance and the same number of layers $p$, \textit{ibm\_kingston} always show higher approximation ratio compared with \textit{ibm\_rensselaer}, which is likely due to its shorter two-qubit gate time and higher gate fidelity. 

\setcounter{table}{0}
\renewcommand{\thetable}{S\arabic{table}} 
\renewcommand{\tablename}{Table}  

\begin{table}[h]

\caption{
Performance of LC-QAOA on unweighted random $3$-regular MaxCut instances with 100 vertices executed on \textit{ibm\_rensselaer} and \textit{ibm\_kingston}. 
\textbf{Instance}: $(N,d,s,u/w)$ specifies the number of vertices $N$, degree $d$ , seed $s$ to generate the instance. $u$ means unweighted and $w$ means weighted. 
\textbf{True MaxCut}: the true MaxCut value of the instance.
\textbf{QPU}: quantum processor used for execution. 
\textbf{$p$}: number of layers. 
\textbf{Mean AR}: mean approximation ratio of the sampled bitstring set (round to two decimal places).
\textbf{Best AR}: best approximation ratio among the sampled bitstring set (round to two decimal places).
\textbf{Num. of it}: number of optimization iterations performed.}
\label{tab:S1}
\begin{ruledtabular}
\begin{tabular}{cccccccc}
Instance & True MaxCut & QPU & $p$ & Mean AR & Best AR & Num. of it \\
\hline
(100,3,8,u)  & 137 & \texttt{ibm\_rensselaer} & 1 & 0.67 & 0.80 & 36 \\
(100,3,12,u) & 135 & \texttt{ibm\_rensselaer} & 1 & 0.69 & 0.81 & 29 \\
(100,3,42,u) & 135 & \texttt{ibm\_rensselaer} & 1 & 0.68 & 0.81 & 31 \\
(100,3,68,u) & 136 & \texttt{ibm\_rensselaer} & 1 & 0.67 & 0.80 & 29 \\
(100,3,75,u) & 137 & \texttt{ibm\_rensselaer} & 1 & 0.67 & 0.80 & 26 \\
(100,3,8,u)  & 137 & \texttt{ibm\_kingston}   & 1 & 0.69 & 0.85 & 38 \\
(100,3,12,u) & 135 & \texttt{ibm\_kingston}   & 1 & 0.70 & 0.84 & 37 \\
(100,3,42,u) & 135 & \texttt{ibm\_kingston}   & 1 & 0.71 & 0.84 & 37 \\
(100,3,68,u) & 136 & \texttt{ibm\_kingston}   & 1 & 0.70 & 0.82 & 37 \\
(100,3,75,u) & 137 & \texttt{ibm\_kingston}   & 1 & 0.69 & 0.80 & 29 \\
(100,3,8,u)  & 137 & \texttt{ibm\_rensselaer} & 2 & 0.68 & 0.81 & 57 \\
(100,3,12,u) & 135 & \texttt{ibm\_rensselaer} & 2 & 0.68 & 0.84 & 55 \\
(100,3,42,u) & 135 & \texttt{ibm\_rensselaer} & 2 & 0.68 & 0.81 & 45 \\
(100,3,68,u) & 136 & \texttt{ibm\_rensselaer} & 2 & 0.69 & 0.87 & 52 \\
(100,3,75,u) & 137 & \texttt{ibm\_rensselaer} & 2 & 0.68 & 0.81 & 44 \\
(100,3,8,u)  & 137 & \texttt{ibm\_kingston}   & 2 & 0.71 & 0.83 & 38 \\
(100,3,12,u) & 135 & \texttt{ibm\_kingston}   & 2 & 0.72 & 0.87 & 43 \\
(100,3,42,u) & 135 & \texttt{ibm\_kingston}   & 2 & 0.74 & 0.87 & 42 \\
(100,3,68,u) & 136 & \texttt{ibm\_kingston}   & 2 & 0.74 & 0.88 & 45 \\
(100,3,75,u) & 137 & \texttt{ibm\_kingston}   & 2 & 0.70 & 0.82 & 44 \\

\end{tabular}
\end{ruledtabular}
\end{table}

\setcounter{figure}{0}
\renewcommand{\thefigure}{S\arabic{figure}} 
\renewcommand{\figurename}{Figure}          

\begin{figure}[h]
\centering
\includegraphics[width=0.6\textwidth]{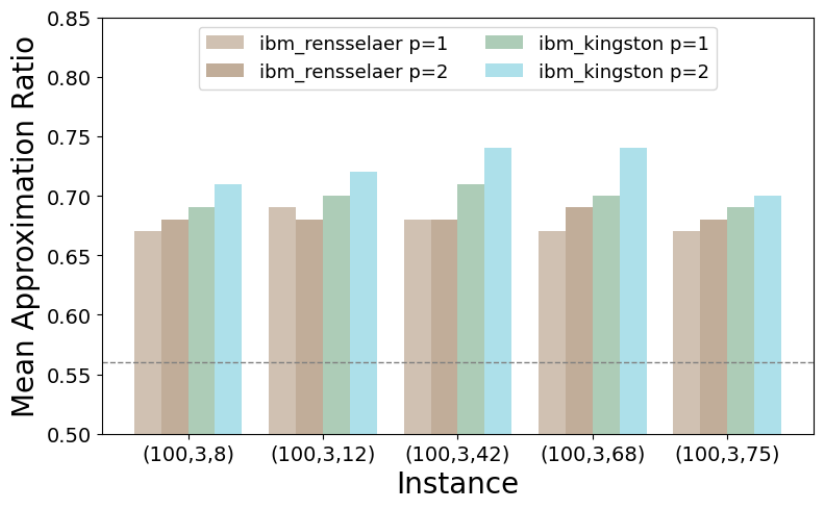}
\caption{
\textbf{Performance of LC-QAOA$_1$ and LC-QAOA$_2$ on random $3$-regular MaxCut instances with 100 vertices.} Horizontal dash line denote mean approximation ratio of unoptimized state. Note that the u indicating unweighted graphs is omitted from the x-axis labels for readability.
}
\end{figure}
\newpage

\subsection{Performance of LC-QAOA$_1$ on random $d$-regular MaxCut instances with 100 vertices}
Here we list instances of solving MaxCut on multiple random $d$-regular graphs using \textit{ibm\_rensselaer} and \textit{ibm\_rensselaer}.
The result shows that as the degree of the random regular graph increases, the mean approximation ratio moderately increases. At the same time, it is worth noticing that the random guess approximation ratio also increases as the degree of the graph increases. Compared with \texttt{ibm\_rensselaer}, \texttt{ibm\_kingston} shows better performance in all cases.

\begin{figure}[h]
\centering
\includegraphics[width=\textwidth]{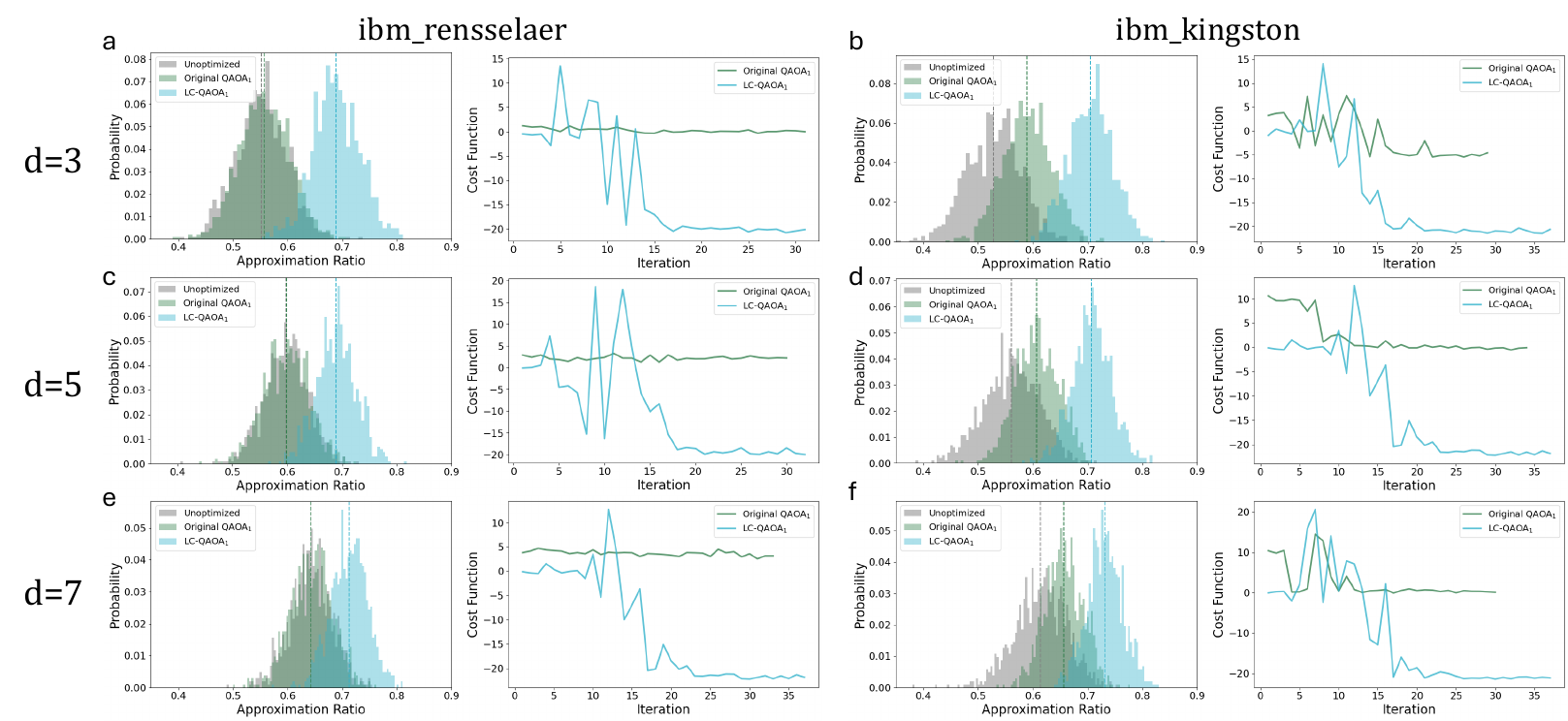}
\caption{
\textbf{Performance of LC-QAOA$_1$ on random $d$-regular MaxCut instances with 100 vertices.}
Approximation ratio probability distribution and cost function evolution of
\textbf{a,} random $3$-regular graph executed by \texttt{ibm\_rensselaer}.
\textbf{b,} random $3$-regular graph executed by \texttt{ibm\_kingston}.
\textbf{c,} random $5$-regular graph executed by \texttt{ibm\_rensselaer}.
\textbf{d,} random $5$-regular graph executed by \texttt{ibm\_kingston}.
\textbf{e,} random $7$-regular graph executed by \texttt{ibm\_rensselaer}.
\textbf{f,} random $7$-regular graph executed by \texttt{ibm\_kingston}.
}
\end{figure}


\subsection{Performance of LC-QAOA on random $3$-regular MaxCut instances with different number of vertices}
Here we show the result of multiple RR3 graphs with different sizes to validate the performance of LC-QAOA ansatz on different sizes of graph. Limited by the length of longest chain in quantum hardware, the maximum number of qubits we can use for this ansatz is 122 on \texttt{ibm\_kingston}. The result reveal that the mean approximation ratio is stable on random $3$-regular MaxCut instance with 40-120 vertices.

\begin{table}[h]
\caption{
Performance of LC-QAOA$_1$ on unweighted random $3$-regular MaxCut instances with multiple number of vertices executed on \textit{ibm\_kingston}. 
\textbf{Instance}: $(N,d,s,u/w)$ specifies the number of vertices $N$, degree $d$ , seed $s$ to generate the instance. $u$ means unweighted and $w$ means weighted. 
\textbf{True MaxCut}: the true MaxCut value of the instance.
\textbf{QPU}: quantum processor used for execution. 
\textbf{$p$}: number of layers. 
\textbf{Mean AR}: mean approximation ratio of the sampled bitstring set (round to two decimal places).
\textbf{Best AR}: best approximation ratio among the sampled bitstring set (round to two decimal places).
\textbf{Num. of it}: number of optimization iterations performed.}
\begin{ruledtabular}
\begin{tabular}{cccccccc}
Instance & True MaxCut & QPU & $p$ & Mean AR & Best AR & Num. of it \\
\hline
(40,3,8,u)   & 55  & \texttt{ibm\_kingston} & 1 & 0.70 & 0.89 & 32 \\
(40,3,12,u)  & 54  & \texttt{ibm\_kingston} & 1 & 0.72 & 0.93 & 31 \\
(40,3,42,u)  & 55  & \texttt{ibm\_kingston} & 1 & 0.70 & 0.87 & 35 \\
(60,3,8,u)   & 83  & \texttt{ibm\_kingston} & 1 & 0.71 & 0.88 & 28 \\
(60,3,12,u)  & 81  & \texttt{ibm\_kingston} & 1 & 0.72 & 0.89 & 34 \\
(60,3,42,u)  & 82  & \texttt{ibm\_kingston} & 1 & 0.72 & 0.87 & 31 \\
(80,3,8,u)   & 109 & \texttt{ibm\_kingston} & 1 & 0.71 & 0.83 & 30 \\
(80,3,12,u)  & 110 & \texttt{ibm\_kingston} & 1 & 0.69 & 0.83 & 29 \\
(80,3,42,u)  & 109 & \texttt{ibm\_kingston} & 1 & 0.71 & 0.86 & 28 \\
(100,3,8,u)  & 109 & \texttt{ibm\_kingston} & 1 & 0.71 & 0.83 & 30 \\
(100,3,12,u) & 135 & \texttt{ibm\_kingston} & 1 & 0.70 & 0.83 & 31 \\
(100,3,42,u) & 135 & \texttt{ibm\_kingston} & 1 & 0.70 & 0.83 & 30 \\
(120,3,8,u)  & 163 & \texttt{ibm\_kingston} & 1 & 0.70 & 0.83 & 30 \\
(120,3,12,u) & 165 & \texttt{ibm\_kingston} & 1 & 0.69 & 0.81 & 34 \\
(120,3,42,u) & 166 & \texttt{ibm\_kingston} & 1 & 0.69 & 0.80 & 31 \\

\end{tabular}
\end{ruledtabular}
\end{table}

\subsection{Performance of LC-QAOA on weighted random $d$-regular graph MaxCut instances}
Here we apply our LC-QAOA to multiple weighted random $d$-regular graph MaxCut instances. The weight of edges are chosen randomly from 0.25, 0.5, 0.75 and 1.

\begin{table}[h]
\caption{
Performance of LC-QAOA$_1$ on weighted random $d$-regular MaxCut instances with multiple number of vertices executed on \textit{ibm\_kingston}. 
\textbf{Instance}: $(N,d,s,u/w)$ specifies the number of vertices $N$, degree $d$ , seed $s$ to generate the instance. $u$ means unweighted and $w$ means weighted. 
\textbf{True MaxCut}: the true MaxCut value of the instance.
\textbf{QPU}: quantum processor used for execution. 
\textbf{$p$}: number of layers. 
\textbf{Mean AR}: mean approximation ratio of the sampled bitstring set (round to two decimal places).
\textbf{Best AR}: best approximation ratio among the sampled bitstring set (round to two decimal places).
\textbf{Num. of it}: number of optimization iterations performed.}
\begin{ruledtabular}
\begin{tabular}{cccccccc}
Instance & True MaxCut & QPU & $p$ & Mean AR & Best AR & Num. of it \\
\hline
(50,6,28,w)   & 78    & ibm\_rensselaer & 1 & 0.71 & 0.85 & 28 \\
(50,7,27,w)   & 90    & ibm\_rensselaer & 1 & 0.70 & 0.91 & 31 \\
(60,5,36,w)   & 83.5  & ibm\_rensselaer & 1 & 0.68 & 0.83 & 28 \\
(70,4,75,w)   & 78    & ibm\_rensselaer & 1 & 0.69 & 0.83 & 32 \\
(80,4,46,w)   & 90.25 & ibm\_rensselaer & 1 & 0.68 & 0.80 & 28 \\
(100,3,42,w)  & 90.75 & ibm\_rensselaer & 1 & 0.67 & 0.82 & 31 \\
\end{tabular}
\end{ruledtabular}
\end{table}

\subsection{Effect of Bit-Flip Post-processing}
Results of bit-flip post-processing on more MaxCut instances are shown here. In all instances, the true MaxCut state is found.

\begin{figure}[h]
\centering
\includegraphics[width=\textwidth]{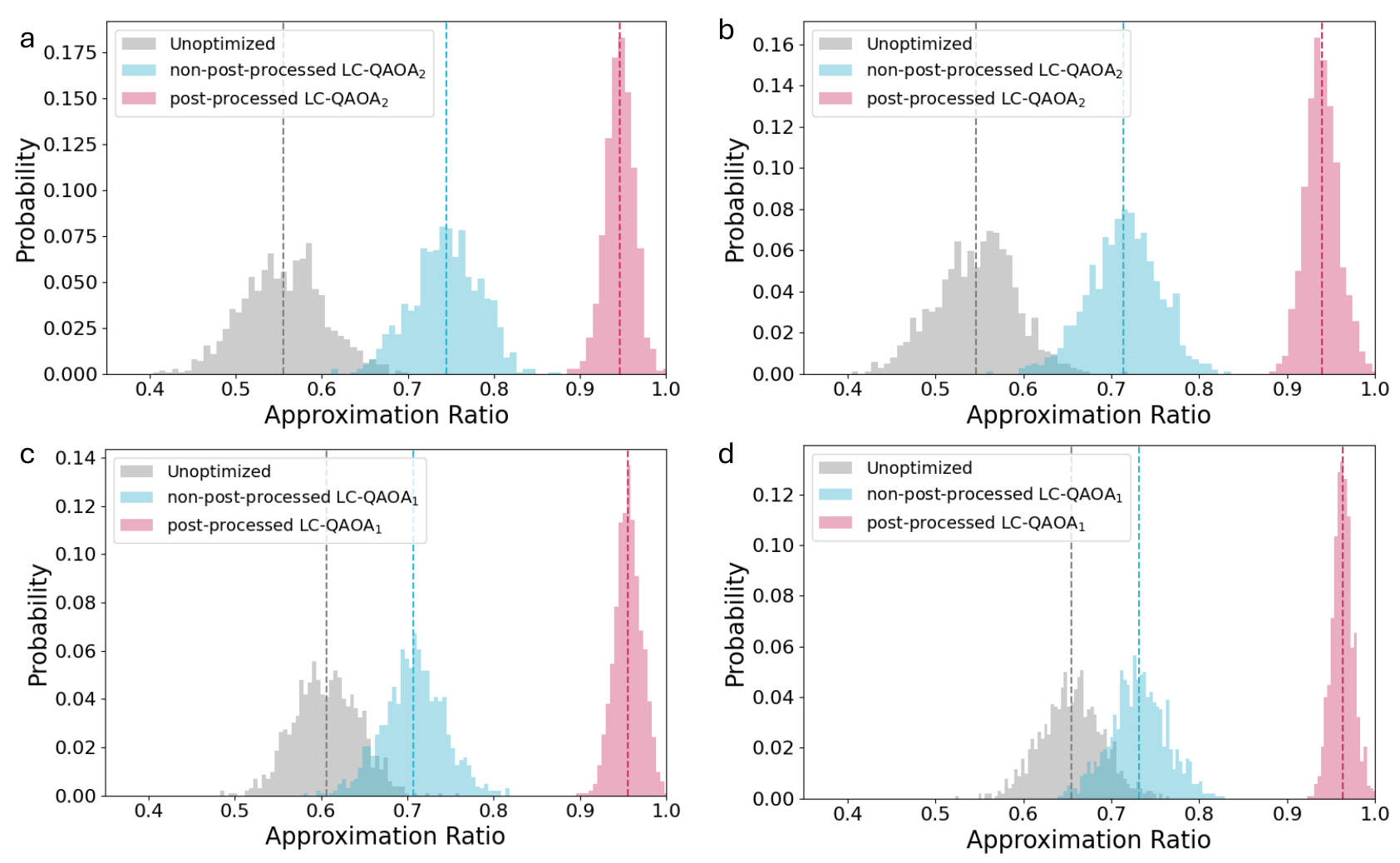}
\caption{
\textbf{Effect of Bit-Flip Post-processing.} 
Probability distribution of approximation ratio of unoptimized state and
\textbf{a,} original and post-processed states of $QAOA_2$ on a random $3$-regular instance with 100 vertices.
\textbf{b,} original and post-processed states of $QAOA_2$ on a random $3$-regular instance with 100 vertices.
\textbf{c,} original and post-processed states of $QAOA_1$ on a random $5$-regular instance with 100 vertices.
\textbf{d,} original and post-processed states of $QAOA_1$ on a random $7$-regular instance with 100 vertices.
}
\end{figure}

\subsection{Effect of chain length on the performance of LC-QAOA ansatz}

The relation between number of vertices under different chain percentages (defined as the ratio between the number of vertices used in the LC ansatz and the total number of vertices in the problem graph) and approximation ratio is illustrated in Fig. S4 (on noiseless \texttt{Aer Simulator}). A Hamiltonian path is first identified using a heuristic depth-first search (DFS) method, after which different percentages of the chain are employed for ansatz construction. The results show that as the chain percentage decreases, the mean approximation ratio gradually declines. When the percentage is reduced to 0\%, the approximation ratio converges to that of a random guess. This trend suggests that even if a Hamiltonian path cannot be identified by the heuristic method, the ansatz remains functional, albeit with reduced performance.

\begin{figure}[h]
\centering
\includegraphics[width=0.6\textwidth]{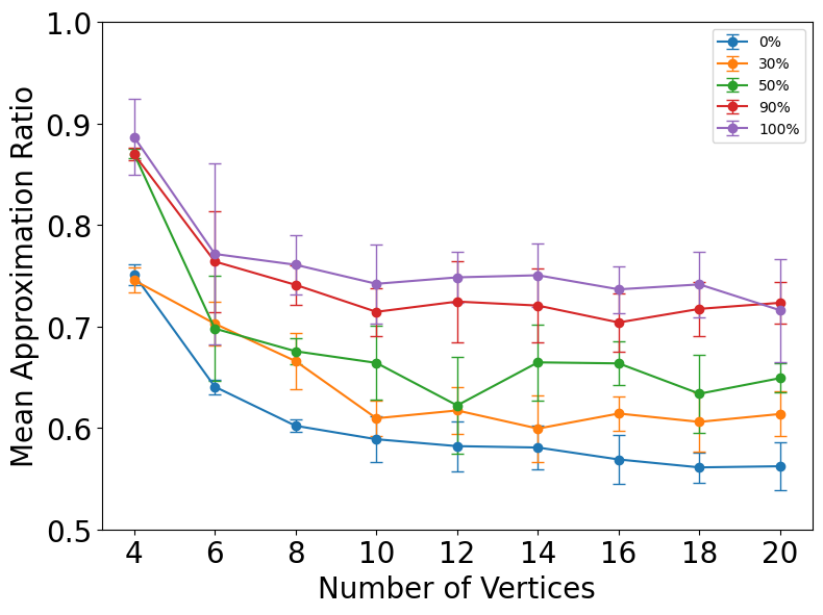}
\caption{
\textbf{Relation between Chain Percentage and Mean Approximation Ratio on Random 3-Regular graph with different number of vertices.} 
Each data point represents the average result over 10 randomly generated random 3-regular graphs instances. Noiseless \texttt{Aer Simulator} is used to run the algorithm. Error bar indicate $\pm\sigma$.
}
\end{figure}

\end{document}